\documentclass{jfm}
\usepackage{graphicx}
\usepackage{epstopdf, epsfig}
\usepackage{amsmath}

\usepackage{graphics}
\usepackage{color, xcolor}
\usepackage{soul}
\soulregister\cite7
\soulregister\citep7
\soulregister\citet7
\soulregister\ref7
\soulregister\pageref7
\shorttitle{Re-picturing viscoelastic drag-reducing turbulence...}
\shortauthor{W. -H. Zhang, H. -N. Zhang, Y. -K. Li, and F. -C. Li}

\title{Re-picturing viscoelastic drag-reducing turbulence by introducing dynamics of elasto-inertial turbulence}

\author{Wenhua Zhang\aff{1,2},
Hongna Zhang\aff{1}
\corresp{\email{hongna@tju.edu.cn}},
Yuke Li\aff{3},
 \and Fengchen Li\aff{1}
 \corresp{\email{lifch@tju.edu.cn}}}

\affiliation{
\aff{1}Key Laboratory of Efficient Utilization of Low and Medium Grade Energy, MOE, School of Mechanical Engineering, Tianjin University, Tianjin 300350, China
\aff{2}Sino-French Institute of Nuclear Engineering and Technology, Sun Yat-sen University, Zhuhai 519082, China
\aff{3}Department of Physics of Complex System, Weizmann Institute of Science, Rehovot 7610001 Israel
}

\begin{document}

\maketitle

\begin{abstract}
Recently, the nature of viscoelastic drag-reducing turbulence (DRT), especially maximum drag reduction (MDR) state, has become a focus of controversy. It has long been regarded as polymers-modulated inertial turbulence (IT), but is challenged by the newly proposed concept of elasto-inertial turbulence (EIT). This study is to re-picture DRT in parallel plane channels by introducing dynamics of EIT based on statistical and budget analysis for a series of flow regimes from the onset of DR to EIT. Energy conversion between velocity fluctuations and polymers as well as polymeric pressure redistribution effect are of particular concern, based on which a new energy self-sustaining process (SSP) of DRT is re-pictured. The numerical results indicate that at low Reynolds number (Re), the flow enters laminar regime before EIT-related SSP is formed with the increase of elasticity, whereas, at moderate Re,  EIT-related SSP can get involved and survive from being relaminarized. This somehow explains the reason why relaminarization is observed for small Re while the flow directly enters MDR and EIT at moderate Re. Moreover, with the proposed energy picture, the newly discovered phenomenon that the streamwise velocity fluctuations lag behind those in wall-normal direction can be well explained. The re-pictured SSP certainly justify that IT nature is gradually replaced by that of EIT in DRT with the increase of elasticity.
\end{abstract}


\section{Introduction}
Adding small quantities of high-molecular weight polymers to Newtonian turbulence can significantly reduce the flow drag. Since its discovery by Toms (1948), extensive researches on its kinematics and dynamics have been carried out (see, e.g., Lumley 1969; White {\&} Mungal 2008; Graham 2014; Xi 2019), and important concomitant phenomena were found. Especially, the MDR phenomenon (Virk 1971), turbulence will not be completely removed, by polymers has long been a challenge for understanding the essence of viscoelastic drag reducing turbulence (DRT). So far, DRT was widely regarded as a perturbation of Newtonian inertial turbulence (IT) by polymers. Starting from this point, various theories of DR induced by polymers including the classical viscous theory (Lumley 1969) and elastic theory (de Gennes 1990) were proposed. MDR state was considered to be a form of ``hibernating'' turbulence, which is inherently part of Newtonian turbulence but becomes unmasked by polymer elasticity (Xi {\&} Graham 2010;  Zhu {\&} Xi 2021). However, the mechanism of DRT has not been unified yet and the proposed theories cannot explain all the accompanying phenomena.

Recently, the knowledge about DRT, especially MDR, has been significantly advanced benefiting from the investigations on elastic turbulence (ET, Groisman {\&} Steinberg 2000) and elasto-inertial turbulence (EIT, Samanta et al. 2013).   Spectra analysis in DRT done by Watanabe {\&} Gotoh 2010 and 2014  implied the possible connection between DRT and ET. Different from pure IT and ET, EIT is induced by elastic nonlinearity and the maintenance of disturbance requires the participation of fluid inertia. One of its distinctive features is the emergence of trains of spanwise oriented vortical structures with alternating sign that appear on elongated sheets of highly stretched polymers (Dubief et al. 2013; Shekar et al. 2019; Sid et al. 2018; Gillissen 2019; Choueiri et al. 2018). Therefore, EIT is essentially regarded as two-dimensional (2D) turbulence (Sid et al. 2018; Gillissen 2019). Today, the connection between DRT and EIT has become a hotspot in the field of visceoasltic turbulence. Samanta et al. (2013) and Dubief et al. (2013) found that the DRT displays features of EIT after the flow enters MDR state. Choueiri et al. (2018) reported that the MDR asymptotic limit can be exceeded by carefully controlling concentration of the dilute polymer solution, and further increasing the polymer concentration will lead the flow to a saturated EIT regime with the flow drag corresponding to that in MDR state. Their observations directly support that the MDR state is essentially an EIT regime, and there exists a ``coexistence phase'' of EIT and IT immediately before MDR.  In our recent work, we developed a characterization method of IT and EIT related dynamics by decomposing the drag coefficients, and proposed a concept that EIT-related dynamics start to play in DRT long before entering MDR and DRT phenomenon is possibly the result of IT and EIT-related dynamics interaction (Zhang et al. 2021b). Those work identifies there does exist a link between DRT and EIT, although it is figured from the perspectives of phenomenological comparisons.

The present paper further attempts to build a mechanistic link by drawing an energy picture for the self-sustaning process (SSP) in DRT considering the role of EIT. Regarding EIT, Dubief et al. (2013) first proposed its SSP cycle, later confirmed by Terrapon et al. 2015, in which they argue that the unstable nature of the nonlinear advection of polymers plays an important role, and once triggered, EIT is self-sustained and fed upon the velocity fluctuations created by the elastic instability. In a recent work (Zhang et al., 2021a), we also pictured the energy transfer process in EIT and argue that rather than the nonlinear advection of polymers, the nonlinear part of elastic shear stress (ESS, the so-called ``stress loss''), originated from the polymers and turbulence interaction, plays a dominant role in forming sheet-like structures and supplying energy to turbulent structures so as to sustain EIT.  The role of inertia seems to lift the polymer sheets towards the channel center, pumping energy from near-wall region to the bulk region. Although its SSP still needs to be further uncovered, it is certain that EIT follows a very different turbulence self-sustaining mechanism from that of IT.

The present paper aims at re-picturing the energy transfer process in DRT by introducing EIT-related dynamics and revealing the potential link between DRT and EIT. It is organized as follows: Section 2 discribes the methods used in this study, including problem formulation and computations details, budget equations as well as the decompositions of ESS and pressure fluctuations; Section 3 presents and analyzes the obtained results including the statistics and the energy and stress budgets, and then propose an energy picture for SSP in DRT; Section 4 gives the conclusions and outlooks.

\section{Methods}
\subsection{Problem formulation and computational details}

This work employs datasets from our recent DNSs of DRT passing a parallel plane channels with the Oldroyd-B model, which have been soundly validated (Zhang et al. 2021a and 2021b). In the three-dimensional channels, \emph{x}, \emph{y} and \emph{z} are the streamwise, wall-normal and spanwise respectively, and the corresponding velocity components are \emph{u}, \emph{v} and \emph{w}, abbreviated as ${u_i}$ (${i= 1,2,3 }$ represents \emph{x}, \emph{y} and \emph{z} directions). For brevity, variables with superscript ``*" are dimensionless variables based on outer scale, and those without superscript are original ones. The outer scale takes the channel half height ${h}$ as the reference length, the bulk mean velocity ${u_\text{b} = \frac{1}{h}\int_0^\delta  {{{\bar u}^{}}d{y^ * }} }$(${\bar u}$ is the local mean streamwise velocity) as the reference velocity, and ${h/u_\text{b}}$ and ${\rho}u_\text{b}^2$  as the reference time and stress, respectively. Neglecting the volume force, the dimensionless governing equations based on the outer scale and Oldroyd-B model are:

\begin{equation}
\frac{{\partial u_i^ * }}{{\partial x_i^ * }} = 0\label{EQ.1},
\end{equation}

\begin{equation}
\frac{{\partial u_i^ * }}{{\partial {t^ * }}} + u_j^ * \frac{{\partial u_i^ * }}{{\partial x_j^ * }} =  - \frac{{\partial {p^ * }}}{{\partial x_i^ * }} + \frac{2}{{{\mathop{\rm Re}\nolimits} }}\frac{\partial }{{\partial x_j^ * }}\left( {\frac{{\partial u_i^ * }}{{\partial x_j^ * }}} \right) + \frac{{\partial \tau _{ij}^ * }}{{\partial x_j^ * }}\label{EQ.2},
\end{equation}

\begin{equation}
\frac{{\partial {c_{ij}}}}{{\partial {t^ * }}} + u_k^ * \frac{{\partial {c_{ij}}}}{{\partial x_k^ * }} = {c_{ik}}\frac{{\partial u_j^ * }}{{\partial x_k^ * }} + {c_{kj}}\frac{{\partial u_i^ * }}{{\partial x_k^ * }} - \frac{1}{{Wi}}\left( {{c_{ij}} - {\delta _{ij}}} \right),\label{EQ.3}
\end{equation}
where, ${t^*}$ is time; ${p^*}$ is pressure; ${\tau _{ij}^ *}$ is the elastic stress, and is calculated by ${\tau^*_{ij} = \frac{\beta(c_{ij}-\delta_{ij})}{\text{ReWi}}}$; ${c _{ij}^ *}$ is the conformation tensor; ${\delta _{ij}^ *}$ is the unit tensor; ${\text{Re}}=\it{2h u_b/\nu}$ is Reynolds number using ${2h}$; ${\text{Wi}}=\it{\lambda u_b/ \delta}$ is Weissenberg number; ${\beta =\eta/\nu}$ is the viscosity contribution ratio; $\eta$ and $\nu$ are contributions of additives and solvent to zero shear viscosity of solution, respectively; $\lambda$ is the relaxation time of polymer solution.

The datasets under a wide range of flow conditions are selected at $\beta=1/9$, Re=2500--20000, and Wi=0--60 (Wi=0 denotes Newtonian IT) which covers a series of flow regime including IT and DRT from the onset of DR to EIT as well as the re-laminarized flow regime by polymers. The numerical details can be found in the supplementary materials or our previous work (Zhang et al. 2021a and 2021b).

\subsection{Budget equations}
To understand the SSP of DRT, necessary budget equations including decompositions of ESS and pressure fluctuations involved are given below.

The Reynolds stress budget equation can be obtained by re-arranging equation (\ref{EQ.2}):
\begin{equation}
{\partial \overline {u_i^{\rm{'*}}u_j^{\rm{{'*}}}} }/{{\partial {t^*}}} ={P_{ij}^{\rm{R}}}{\rm{ + }}{\Phi _{ij}^{}}+{\varepsilon _{ij}^{\rm{R}}}-{G_{ij}^{}}+{T_{ij}^{\rm{R}}}
,\label{EQ.4}
\end{equation}
where, ${P_{ij}^{\rm{R}}=  { - \left( {\overline {u_i^ {'*} u_k^ {'*} } \frac{{\partial \bar u_j^ {*} }}{{\partial x_k^ {*} }} + \overline {u_j^ {'*} u_k^ {'*} } \frac{{\partial \bar u_i^ {*} }}{{\partial x_k^ {*} }}} \right)}}$ is production rate; ${\Phi _{ij}^{}={\overline {{{p}^ {'*} }\left( {\frac{{\partial u_j^ {'*} }}{{\partial x_i^ {*} }} + \frac{{\partial u_i^ {'*} }}{{\partial x_j^ {*} }}} \right)} }}$ is pressure-strain term; ${\varepsilon _{ij}^{}={ - \frac{4}{{{\mathop{\rm Re}\nolimits} }}\overline {\frac{{\partial u_i^ {'*} }}{{\partial x_k^ {*} }}\frac{{\partial u_j^ {'*} }}{{\partial x_k^ {*} }}} }}$ is dissipation rate; ${G_{ij}={\left( {\overline {\tau _{jk}^{\rm{{'*}}}\frac{{\partial u_i^{\rm{{'*}}}}}{{\partial x_k^{\rm{{*}}}}}}  + \overline {\tau _{ik}^{\rm{{'*}}}\frac{{\partial u_j^{\rm{{'*}}}}}{{\partial x_k^{\rm{{*}}}}}} } \right)}}$ is elastic stress-strain term; $\frac{\partial }{\partial y_{{}}^{*}}\left( -\partial \overline{{u}_{i}^{'*}{u}_{j}^{'*}{v}_{{}}^{'*}}+\partial \overline{{\tau }_{j2}^{'*}{u}_{i}^{'*}}+\partial \overline{{\tau }_{i2}^{'*}{u}_{j}^{'*}}+\frac{2}{\operatorname{Re}}\frac{\partial }{\partial y_{{}}^{*}}\overline{{u}_{i}^{'*}{u}_{j}^{'*}} \right)-\left( \overline{\frac{\partial {{{{p}}}^{'*}}{u}_{j}^{'*}}{\partial x_{i}^{*}}}+\overline{\frac{\partial {{{{p}}}^{'*}}{u}_{i}^{'*}}{\partial x_{j}^{*}}} \right)$ is transport rate.

Pressure redistribution term $\Phi_{ij}$ is an important link in turbulent SSP, responsible for the generation of TKE in the wall-normal direction. Following Ptasinski et al., (2003) and Terrapon et al. (2015), the fluctuating pressure $p'$ can be decomposed into rapid, slow and polymer contributions to investigate the modulation effect of polymers, as  ${{{p'}^ {*} }\left( x \right){\rm{ = }}{p'}_R^ {*} \left( x \right) +{p'}_s^ {*} \left( x \right) +{p'}_p^ {*} \left( x \right)}$, where:
\begin{equation}
{\frac{{{\partial ^2}p_{\rm{R}}^ {'*} }}{{{\partial ^2}x_i^{ * 2}}} =  - 2\frac{{d{{\bar u}^*}}}{{d{y^ * }}}\frac{{\partial v'}}{{\partial {x^ * }}}},   {\frac{{{\partial ^2}p_{\rm{S}}^ {'*} }}{{{\partial ^2}x_i^{ * 2}}} =  - \frac{{\partial u_i^ {'*} }}{{\partial x_j^ * }}\frac{{\partial u_j^ {'*} }}{{\partial x_i^ * }} + \frac{{{d^2}\overline {{{v}^{ '* 2}}} }}{{d{y^{ * 2}}}}},  {\frac{{{\partial ^2}p_{\rm{P}}^ {'*} }}{{{\partial ^2}x_i^{ * 2}}} = \frac{{{\partial ^2}\tau _{ij}^ {'*} }}{{\partial x_i^ * \partial x_j^ * }}}\label{EQ.5},
\end{equation}
and the first two are identified as inertial contribution.

The elastic stress budget equation can be obtained by arranging equation (\ref{EQ.3}):

\begin{equation}
{{\partial \bar \tau _{ij}^ {*} }}/{{\partial {t^ {*} }}} = {P_{ij}^{\rm{E}}} +{G_{ij}^{}}+{T_{ij}^{\rm{E}}}{\rm{ + }}
{V_{ij}^{}}+{\varepsilon _{ij}^{\rm{E}}}\label{EQ.6},
\end{equation}
where, ${P_{ij}^{\rm{E}}}={\left( {\bar \tau _{ik}^ {*} \frac{{\partial \bar u_j^ {*} }}{{\partial x_k^ {*} }} + \bar \tau _{kj}^ {*} \frac{{\partial \bar u_i^ {*} }}{{\partial x_k^ {*} }}} \right)}$ is production rate;  ${G_{ij}^{}}$ is elastic stress-strain term; ${T_{ij}^{\rm{E}}}={ - \overline {\frac{{\partial v^ {'*} \tau _{ij}^ {'*} }}{{\partial y^ {*} }}} }$ is turbulent transport rate;  ${V_{ij}^ * ={\frac{{2\beta }}{{{\mathop{\rm Re}\nolimits} Wi}}\left( {\frac{{\partial \bar u_j^ {*} }}{{\partial x_i^ {*} }} + \frac{{\partial \bar u_i^ {*} }}{{\partial x_j^ {*} }}} \right)}}$ is viscous effect term;  ${\varepsilon _{ij}^{\rm{E}}={ - \frac{1}{{Wi}}\bar \tau _{ij}^ {*} }}$ is elastic disspation rate.

As the excitation of EIT is independent of the viscous effect term in (\ref{EQ.6}) which is a linear term, the mean ESS can be decomposed into linear part $\bar \tau _{E1}^ *$ and nonlinear part $\bar \tau _{E2}^ *$ (see Zhang et al. 2021a and 2021b for more information):
\begin{equation}
\bar \tau _E^ * {\rm{ = }}{\bar \tau _{E1}^ * } + {\bar \tau _{E2}^ * }\label{EQ.7},
\end{equation}
where, ${\bar \tau _{E1}^*={\frac{\beta }{{{\mathop{\rm Re}\nolimits} }}\frac{{\partial \bar u_{}^ * }}{{\partial y_{}^ * }}}}$ ; ${\bar \tau _{E2}^*={Wi\left( {\bar \tau _{yy}^ * \frac{{\partial \bar u_{}^ * }}{{\partial y_{}^ * }}{\rm{ + }}\left( {\overline {\tau _{1k}^{\rm{ '* }}\frac{{\partial v^ {'*} }}{{\partial x_k^ * }}}  + \overline {\tau _{k2}^ {'*} \frac{{\partial u_{}^ {'*} }}{{\partial x_k^ * }}} } \right) - \overline {\frac{{\partial u_k^{'*} \tau _E^ {'*} }}{{\partial x_k^ * }}} } \right)}}$ is caused by the interaction between polymers and turbulence, that is, the so-called "stress loss".
\section{Results and analysis}

\subsection{Statistical properties}

At first, we present some results for the flow statistics. Figure 1 illustrates friction coefficient $f$, global TKE production rate $P_{\rm k}$ by RSS, global TEE production rate $P_{\rm e}$ by NESS $\tau_{\rm E2}$ as well as the global energy transfer rate $G^{*}$ between EE and TKE. The datasets qualitatively reproduces the interesting observations in Choueiri's experiments, which can be read from the drag coefficients $f$ as shown in Figure 1a. At a medium Re (e.g., 6000 or 20000), the viscoelastic turbulence experiences the regimes of the onset of DR, LDR, HDR, MDR and EIT with the increase of Wi. Correspondingly, evolutions of ${P_{\rm{k}}^* }$, ${P_{\rm{e}}^ * }$ and ${G_{}^ * }$ share similar trends for moderate Re. Since the onset of DR, ${P_{\rm{k}}^ * }$ starts to decrease, while ${P_{\rm{e}}^ * }$ starts to increase. When the flow enters MDR and EIT regime, ${P_{\rm{k}}^* }$ becomes negligible, whereas  ${P_{\rm{e}}^ * }$ becomes dominant and saturated. During this process, $G^*$ firstly decreases to a valley and then starts to increase again with Wi. More strikingly, the value of $G^ *$ is negative below the critical Wi of MDR indicating globally the energy is transferred from TKE to TEE, but positive above the critical Wi of MDR indicating the globally inversed energy transfer. Moreover, the evolution of $G^*$ after the valley point for two Re collapse to each other showing a power$-$law relationship with Wi implying a scaling for $G^*$ independent of Re, which needs further investigations in the future. At a small Re (e.g., 2500), it experiences the regimes of the onset of DR, LDR, HDR, laminar flow regime and EIT ($Wi>15$). The existence of the laminarized flow regime for moderate Wi ($\approx 6-15$ in the current work) which distinguishes the origins of DRT in two different stages. Evolutions of ${P_{\rm{k}}^ * }$, ${P_{\rm{e}}^ * }$ and ${G_{}^ * }$ show a different behavior for low Re (e.g., 2500) comparing with that for moderate Re. Before relaminarization, ${P_{\rm{k}}^ * }$ is dominant and the value of $G^*$ is always negative indicating the turbulence is always IT dominated which is inhibited by polymers; when the flow enters laminar regime for Wi $\approx 6-15$, ${P_{\rm{k}}^ * }$, ${P_{\rm{e}}^ * }$ and ${G_{}^ * }$ become eliminated in laminar regime. In contrast, during the process of entering EIT, ${P_{\rm{e}}^ *}$ is dominant and the value of $G^*$ is positive for $Wi>15$ indicating the turbulence is sustained by polymers. The above results demonstrate a global picture for the elastic effect on the viscoelastic turbulence: (1) at small Wi (LDR for medium Re or before relaminarization for small Re), the flow is IT-dominated and in these cases the traditional TKE production by Reynolds stress is suppressed by polymers and thus SSP of IT is modulated; (2) increasing Wi above a critical $Wi_{\rm c}$, a new TKE generation term induced by elastic effect appears, which grows with Wi and changes the SSP cycle of DRT; (3) further increase Wi, the turbulence is sustained from by $P_{\rm k}^*$ to $P_{\rm e}^*$, and the nature of turbulence changes from polymer-modulated IT to EIT.

\begin{figure}
\centering
\includegraphics[width=0.333\textwidth]{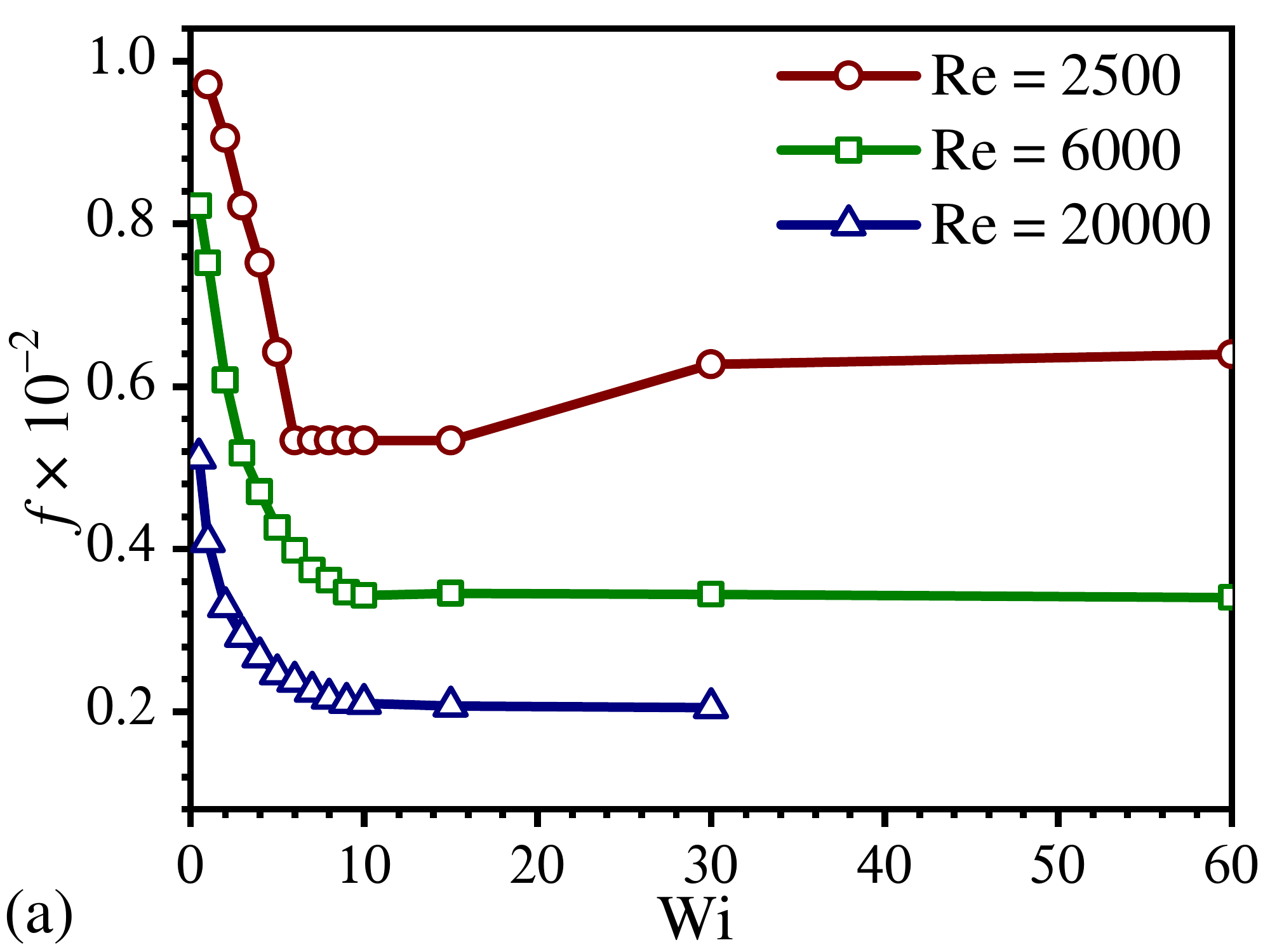}\includegraphics[width=0.333\textwidth]{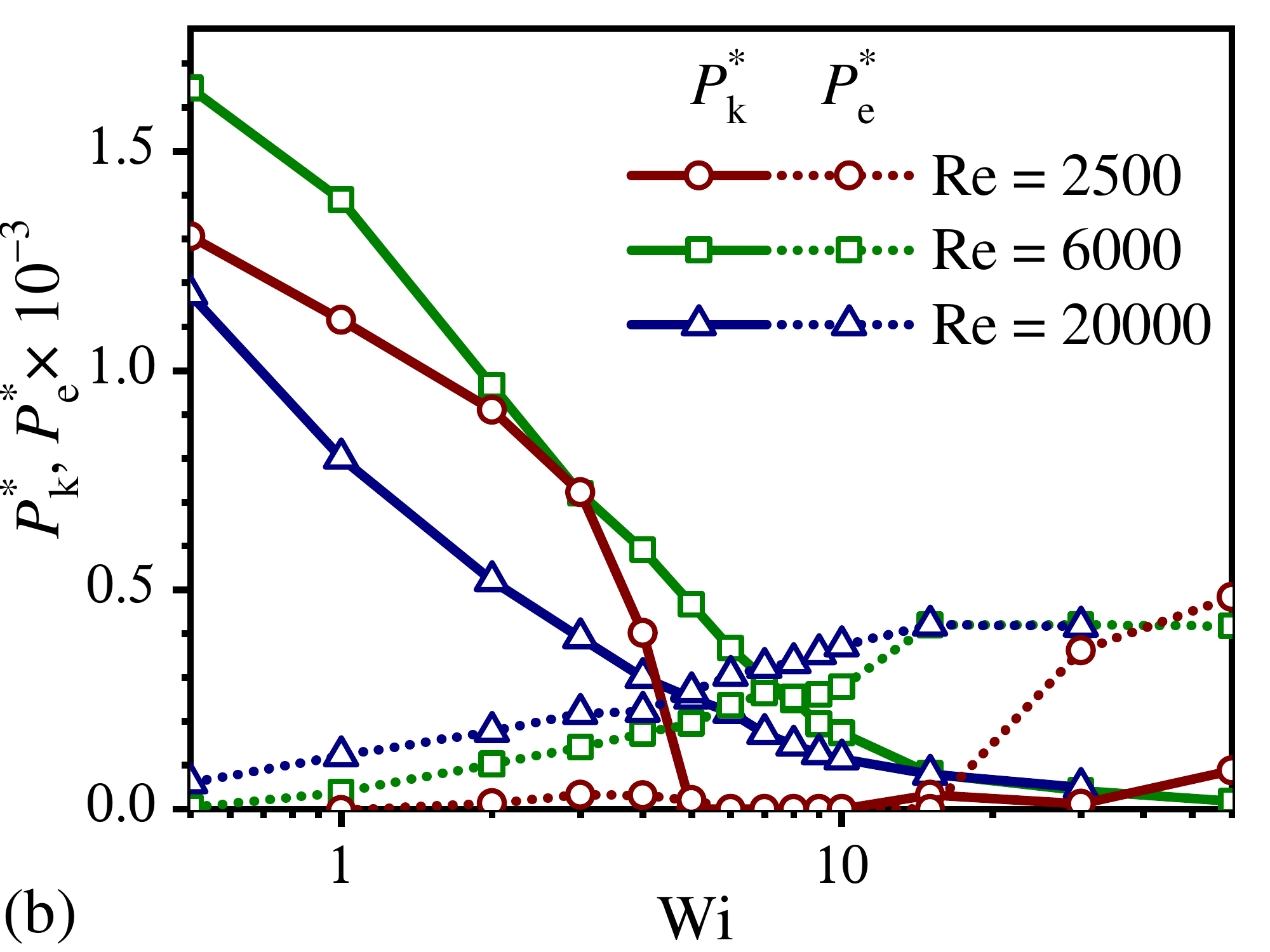}\includegraphics[width=0.333\textwidth]{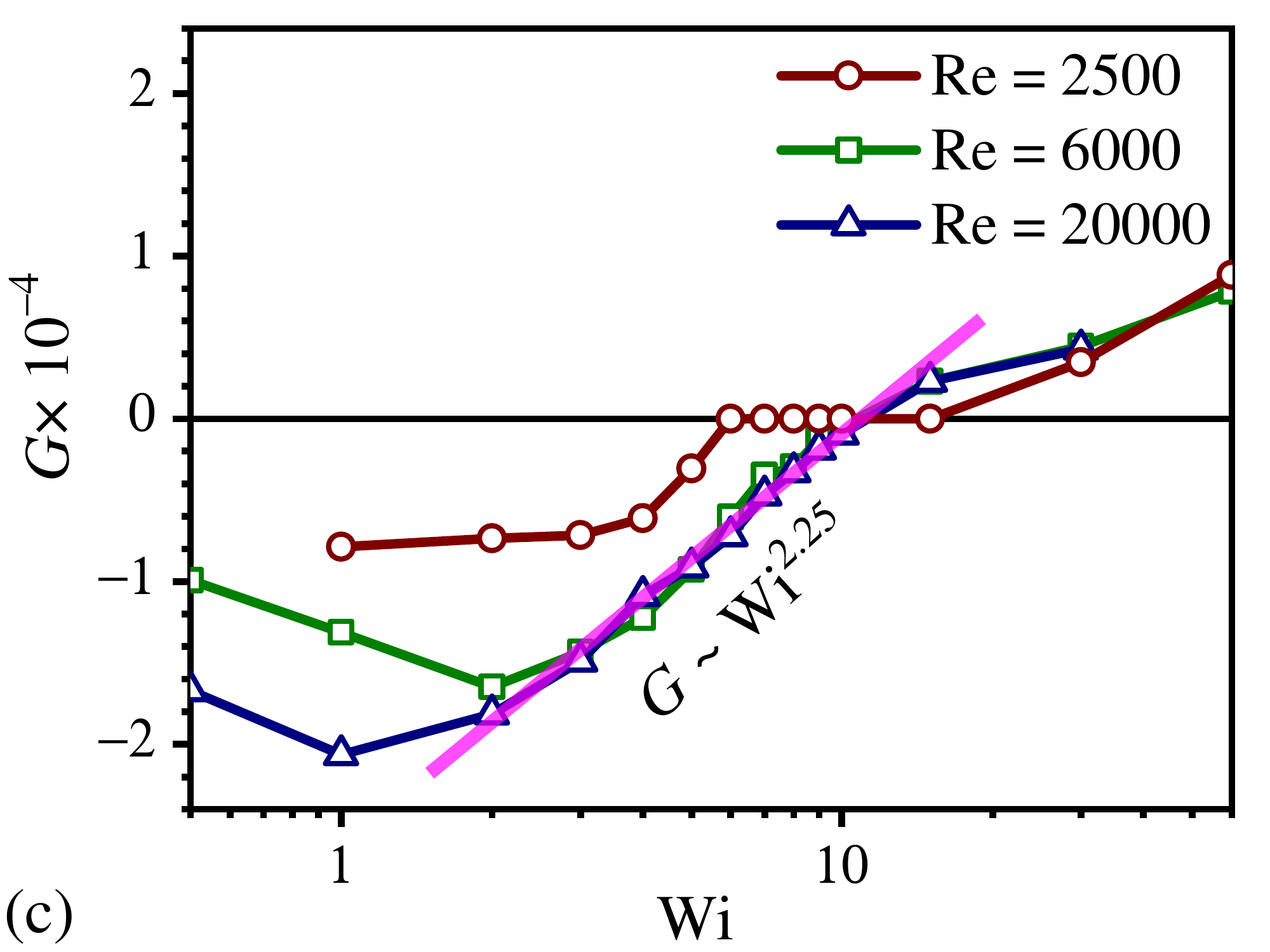}
\caption{\label{fig1} Evolution of statistics with Wi: (a) friction coefficient $f$;  (b) global TKE production ${P_{\rm{k}}^ * {\rm{ = }} - \int_0^2 {\overline {{{u'}^ {*} }{{v'}^ {*} }} \frac{{\partial {{\bar u}^ * }}}{{\partial {y^ * }}}d{y^ * }} }$ and EE production rate ${P_{\rm{e}}^ * {\rm{ = }} - \int_0^2 {\bar \tau _{E2}^{\rm{ + }}\frac{{\partial {{\bar u}^ * }}}{{\partial {y^ * }}}d{y^ * }} }$;  (c) global energy transformation rate ${G^* {\rm{ = }}\int_0^2 {\overline {\tau _{ij}^{'*}\frac{{\partial u_i^{'*}}}{{\partial x_j^*}}} d{y^ * }} }$. $G^* >0$ represents the energy is globally transferred from polymers to the flow structures, otherwise from the flow structures to the polymers.}
\end{figure}

The characteristics of Reynolds stress attract the most attention in the analysis of IT and DRT. However, although there exist obvious $u'$ and $v'$ (especially $u'$) in EIT, $\overline{{{u'}^ {*}}{{v'}^{*}}}$ is extremely small, implying somehow the decoupling of $u'$ and $v'$. From our recent demonstrated SSP cycle of EIT (Zhang et al. 2021b), the TKE production in different directions and the embedded TKE transfer amongst them clearly differ from that of IT, which is possible to delay or decouple the generation of $u'$ from $v'$.  Figure 2 presents the relative cross-correlation $C_{uv}(\xi)$ between $u'$ and $v'$ over $z$ and $t$ where $C_{uv}(\xi)={\left\langle {u'\left( {{\bf{x}},t} \right)v'\left( {{\bf{x}} + {\bf{r}},t} \right)} \right\rangle /\left\langle {u'\left( {{\bf{x}},t} \right)v'\left( {{\bf{x}},t} \right)} \right\rangle }$. Strikingly, unlike that in IT, there exists an obvious lag effect between $u'$ and $v'$ in HDR and EIT: $u'$ lags behind $v'$ ($v'$ induces the generation of $u'$ in other words). Moreover, in EIT, the lagging distance $\xi_{\rm{max}}$ can reach several times of the half channel height $h$. For example, at Re = 1000, Wi = 60, $\xi$ can reach more than $5h$. In addition, $\xi_{\rm{max}}$ gradually increases with an approximately linear relationship with Wi and $\xi_{\rm{max}}$ also increases with Re due to the inertia enhancement. The above results once again imply a different picture of TKE transfer process among velocity components in IT and DRT.

\begin{figure}
\centering
\includegraphics[width=0.29\textwidth]{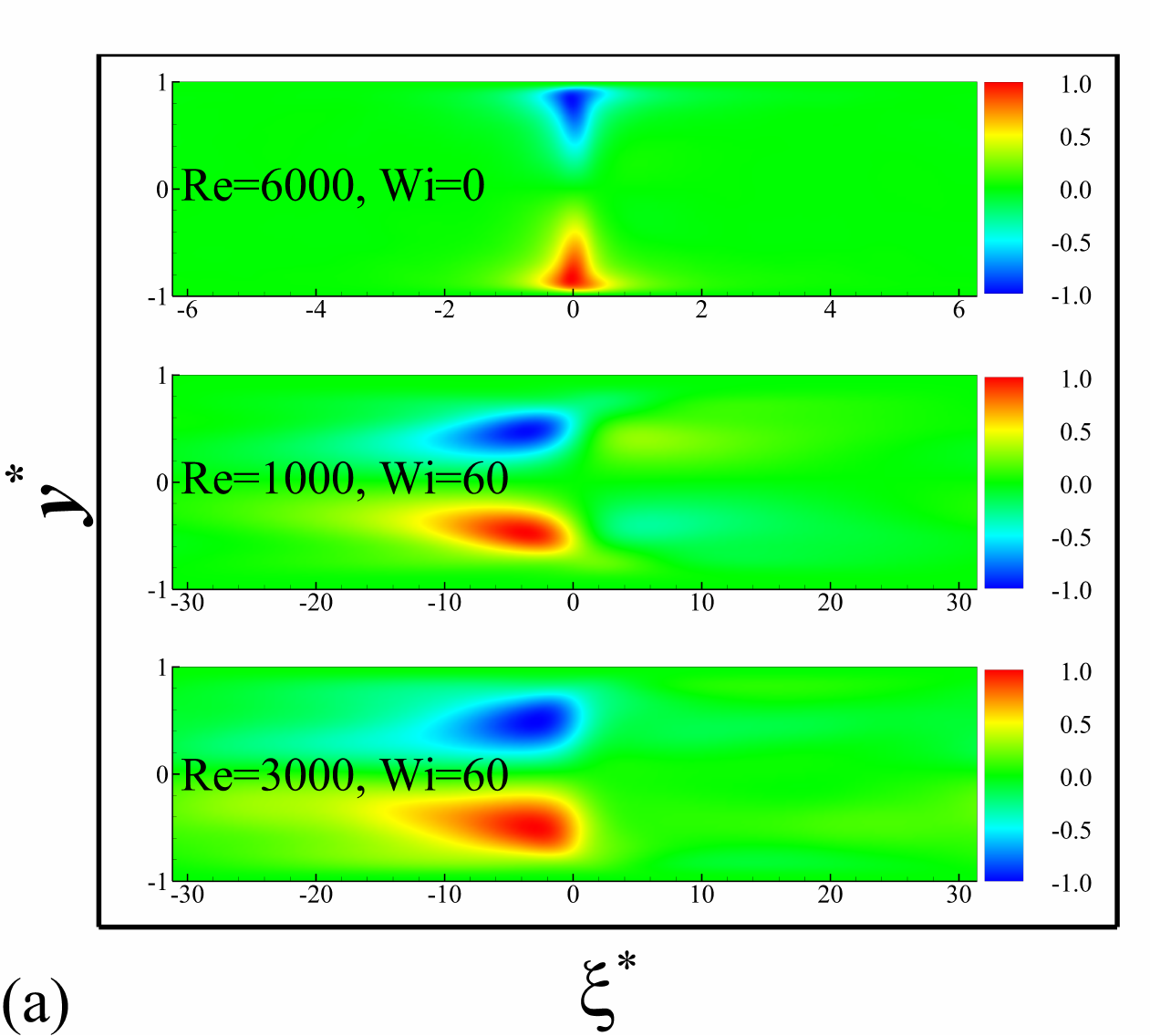}\includegraphics[width=0.4\textwidth]{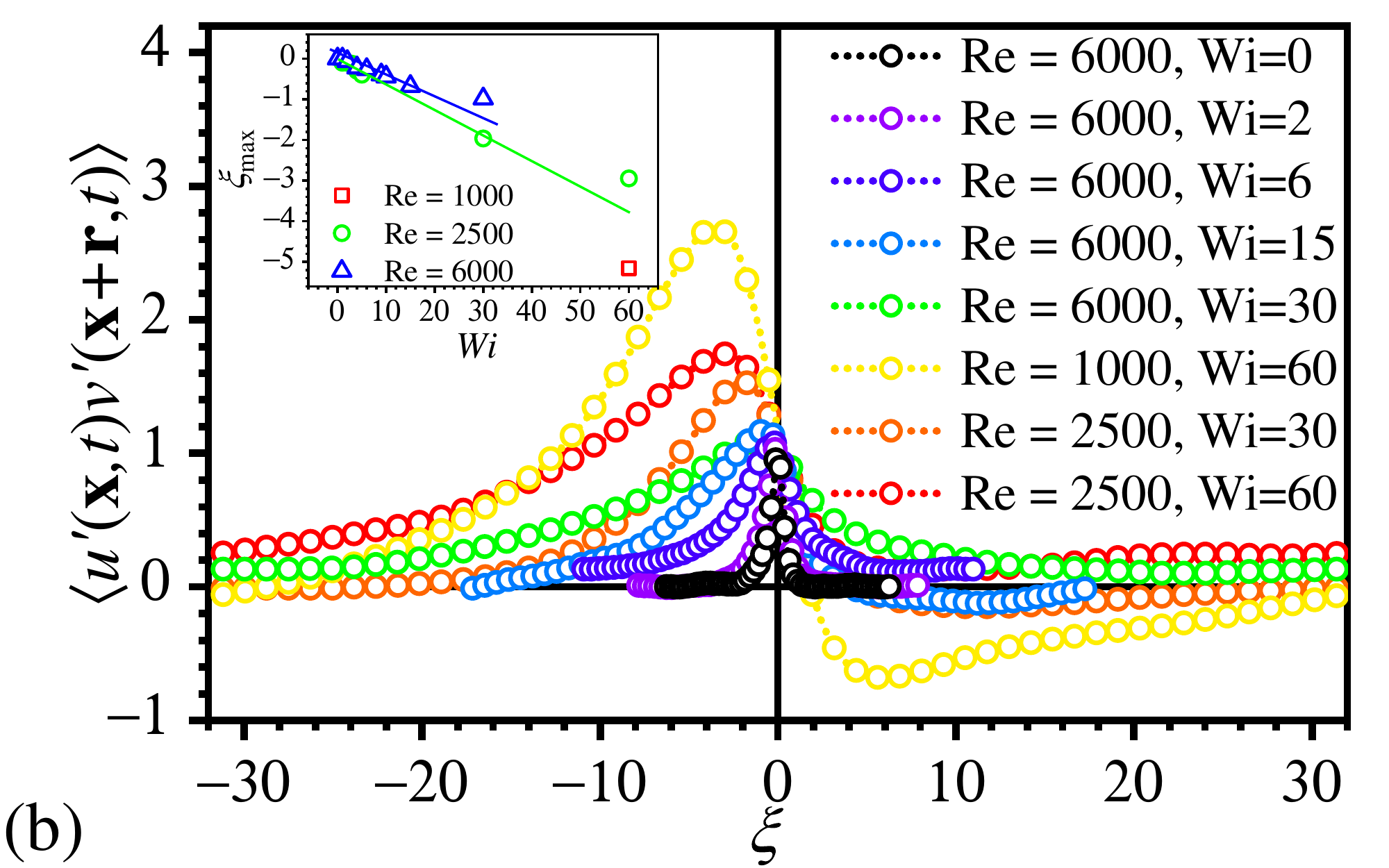}
\caption{\label{fig2} Distributions of $C_{uv}(\xi)$: (a) in $\xi-y$ plane, (b) at $y^* \approx0.5$. Inset: $\xi_{\rm{max}}$ vs. Wi at different Re. $\xi_{\rm{max}}$ corresponds to the $\xi$ where $C_{uv}(\xi)$ reaches the maximum.}
\end{figure}

\subsection{Budget analysis}
This section re-evaluate the energy transfer process in DRT by budget analysis of TKE, EE and stresses among different components as illustrated in Figure 3 considering whether the SSP cycle of EIT is formed.
As well-known, in IT, RSS absorbs energy from the mean motion and produces streamwise TKE, which is transformed into wall-normal TKE through fast and slow pressure redistribution, and regenerates RSS driven by wall-normal velocity fluctuations. The above process forms a cycle to maintain the turbulence against dissipation. As demonstrated in Zhang et al. (2021b), when the flow enters EIT, it follows a very different SSP cycle comparing with that in IT.  The most striking differences lie in:  (1) rather than by RSS, NESS absorbs energy from the mean motion to sustain the turbulence; (2) the conversion of wall-normal TKE into EE and streamwise EE into TKE are added; (3) the interaction between polymers and turbulence suppress both RSS and NESS. In addition, polymers also contributes to modify the energy redistribution term in EIT which affects the generation of $\overline{{{v'}^ {*}}{{v'}^{*}}}$ in the SSP cycle of EIT (Terrapon et al., 2015). EIT-related SSP cycle is possibly formed and survives from being disspated once the above features appear. In the following, special attention will be paid to the performance of terms related with these features.

Common with the earlier studies (Xi, 2019), $P_{xx}^{\rm{R}}$ and $\epsilon_{xx}^{\rm{R}}$ gradually decrease with the increase of Wi, indicating the suppression of elasticity on the IT-related SSP in DRT. Moreover, it is observed that (1) at a medium Re (e.g., 6000),  $P_{xx}^{\rm{E}}$ continuously grows with Wi and exceeds $P_{xx}^{\rm{R}}$ at $Wi\approx9$ in HDR stage, indicating that for cases of $\rm{Wi} > 9$, more energy is transferred from the mean flow to the polymers by NESS than that to the turbulent structures and the turbulence is dominated by NESS; (2) at a small Re (e.g., 2500), the flow is dominated by $P_{xx}^{\rm{R}}$ before the relaminarization while by $P_{xx}^{\rm{E}}$ when EIT is excited. Interesting to notice from profiles of $G_{xx}$ is that it experiences a morphological change with the increase of Wi. For a small Wi (e.g., at Re=6000, $Wi < 3$ and Re=2500, $Wi < 5$ before the re-laminarization), the energy is transferred from $\overline{{{u'}^ {*}}{{u'}^{*}}}$ to the polymers in the bulk region, which increases with Wi and introduces additional energy dissipation. This effect modulates the IT-dynamics related SSP. For moderate and large Wi (e.g., at Re=6000, $Wi > 3$ and Re=2500, $Wi > 15$ after the emergence of EIT regime), the energy transfer direction turns over with the increase of Wi. The negative energy transfer region (from the polymers to $\overline{{{u'}^ {*}}{{u'}^{*}}}$) shifted from the near-wall region to the bulk region, and enlarges with Wi. Meanwhile, the positive energy transfer region (from $\overline{{{u'}^ {*}}{{u'}^{*}}}$ to the polymers) shifted from the bulk region to the near-wall region, and shrinks with the Wi. Notable is that the region where polymers absorb energy shifted upward with Wi, which is probably originated from the turbulence streaks lifting as reported in Zhang et al. (2021b).  The core area is mainly characterized by the energy transfer of elastic energy to $\overline{{{u'}^ {*}}{{u'}^{*}}}$, which indicates that EIT dynamics can be self-sustained, i.e., EIT-related SSP involves. With the increase of Wi, more elastic energy is transferred from polymers to TKE until it completely develops into EIT.

\begin{figure}
\centering
\includegraphics[width=1\textwidth]{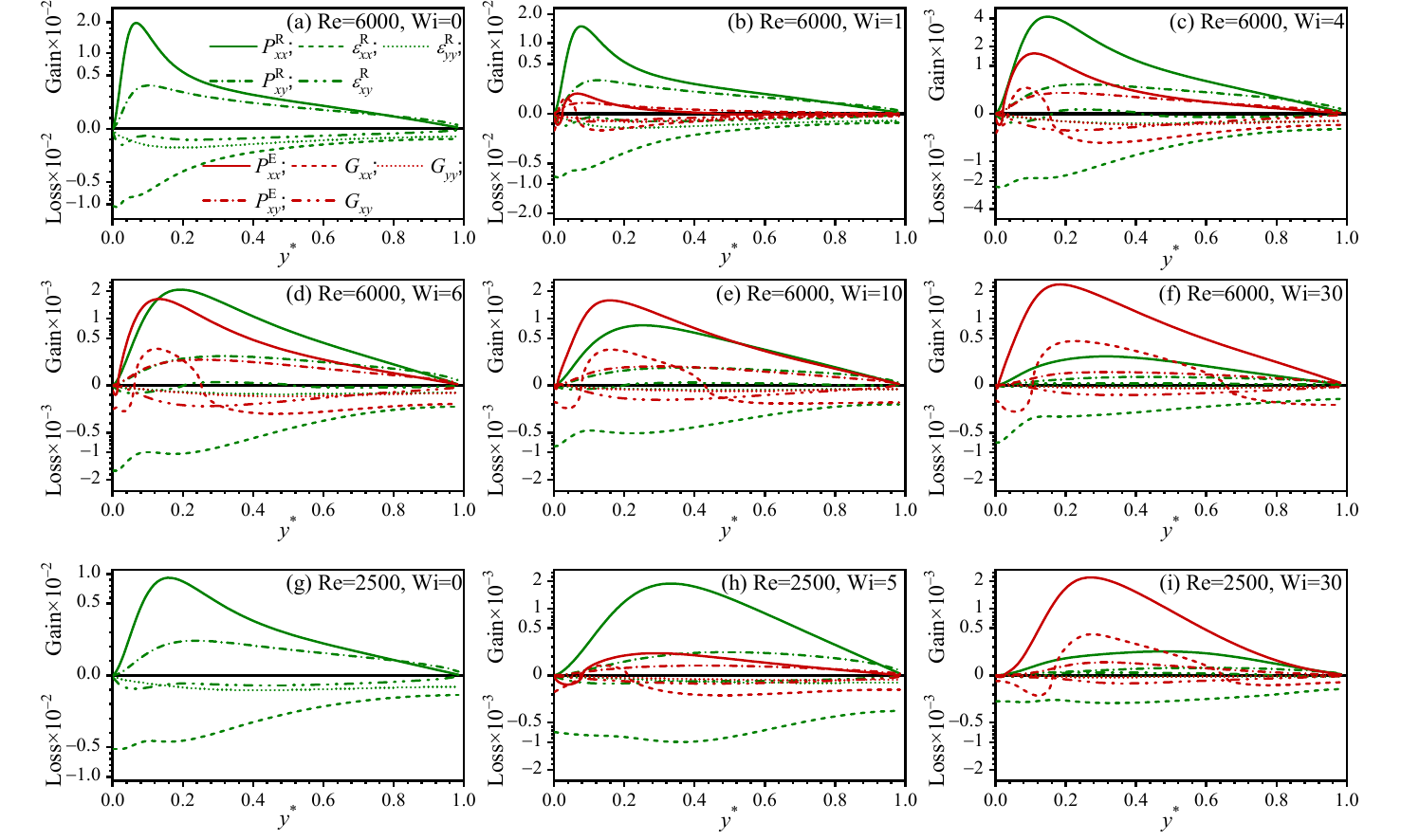}
\caption{\label{fig3} Budgets analysis for $\overline{{{u'}^ {*}}{{u'}^{*}}}$, $\overline{{{v'}^ {*}}{{v'}^{*}}}$, $\overline{{{u'}^ {*}}{{v'}^{*}}}$, $\overline{\tau_{xx}^*}$ and $\overline{\tau_{xy}^*}$ at: (a) Re = 6000, Wi = 0 (Newtonian IT); (b) Re = 6000, Wi = 1; (c) Re = 6000, Wi = 4; (d) Re = 6000, Wi = 6; (e) Re = 6000, Wi = 10 (MDR); (f) Re = 6000, Wi = 30 (EIT); (g) Re = 2500, Wi = 0 (Newtonian IT); (h) Re = 2500, Wi = 5 (before relaminarization); (i) Re = 2500, Wi = 30 (EIT).}
\end{figure}

Comparing with $G_{xx}$, the energy transfer rate $G_{yy}$ between $\overline{{{v'}^ {*}}{{v'}^{*}}}$ and EE is negative for all the covered flow regimes, indicating that the polymers always suppress the normal-wise turbulence. With the increase of Wi, as $\overline{{{v'}^ {*}}{{v'}^{*}}}$ is suppressed by polymers, less energy is transferred from $\overline{{{v'}^ {*}}{{v'}^{*}}}$ to polymers. Different from  $\overline{{{u'}^ {*}}{{u'}^{*}}}$, the generation of $\overline{{{v'}^ {*}}{{v'}^{*}}}$ is from pressure redistribution. Figures 4 shows the evolutions of rapid, slow and polymeric pressure redistribution terms under different flow conditions. In IT both $\phi_{xx}^{\rm{R}}$ and $\phi_{xx}^{\rm{S}}$ behave negative in the turbulence core area, i.e., TKE is redistributed to other directions from $\overline{{{u'}^ {*}}{{u'}^{*}}}$ through pressure fluctuations; the behaviors of both $\phi_{yy}^{\rm{R}}$ and $\phi_{yy}^{\rm{S}}$ are opposite with that of $\phi_{xx}^{\rm{R}}$ and $\phi_{xx}^{\rm{S}}$: near the wall (at $y^* < 0.1$), $\phi_{yy}$ is negative, indicating the energy is redistributed to that in spanwise forming turbulent streaks and in the core area (at $y^* > 0.1$),  $\phi_{yy}$ is positive, indicating the energy is redistributed to that in the wall-normal direction and forms $\overline{{{v'}^ {*}}{{v'}^{*}}}$. In DRT, for small Wi when the turbulence is IT dominated, both $\phi^{\rm{R}}$ and $\phi^{\rm{S}}$ are far larger than $\phi^{\rm{P}}$, and dominate pressure redistribution. With the increase of Wi, $\phi^{\rm{P}}$ is gradually excited on one hand and $\phi^{\rm{R}}$ and $\phi^{\rm{S}}$ are dampened on the other hand. When Wi is large enough, $\phi^{\rm{P}}$ catches up with those two terms, and can affect the pressure redistribution. Strikingly, the evolutions of $\phi^{\rm{P}}$ demonstrate a similar trend with that of $G_{xx}$ and $P^{\rm E}_{xy}$ with the increase of Wi. For cases of small Wi, e.g., $Wi < 3$ at Re = 6000 and $Wi < 5$ at Re = 2500,  $\phi_{yy}^{\rm{P}}$ is small and negative, i.e., $\overline{{{v'}^ {*}}{{v'}^{*}}}$ is transferred to other directions through the polymeric pressure term. However, for large Wi, e.g., $Wi > 3$ at Re = 6000 and $Wi > 30$ at Re = 2500, there is an obvious positive polymer effect near the wall, i.e., TKE in other directions is transferred to the wall-normal direction through the polymer pressure effect. Polymers play a gain effect on the normal-wise TKE generation which sustains the generation of  $\overline{{{v'}^ {*}}{{v'}^{*}}}$ and turbulent SSP. Therefore, the existence of elastic effect introduces a new pressure redistribution mechanism in DRT. It plays a modulation pressure redistribution effect when Wi is small, but can reshape the pressure redistribution effect when Wi is large enough, making the correlation between ${u'}^{*}$ and ${v'}^{*}$ in DRT more complex. When entering EIT, the characteristics of pressure redistribution effect are obviously different from that in IT, and the polymer contribution plays a role in sustaining ${v'}^{*}$ generation.

From the above budget analysis, there is a distinct SSP from IT in viscoelastic DRT. On one hand, the existence of elastic effect suppresses the self-sustaining mechanism related to inertial turbulence. On the other hand, when Wi is large enough, a cycle of EIT-related SSP is introduced, which finally replaces that of IT.
\begin{figure}
\centering
\includegraphics[width=1\textwidth]{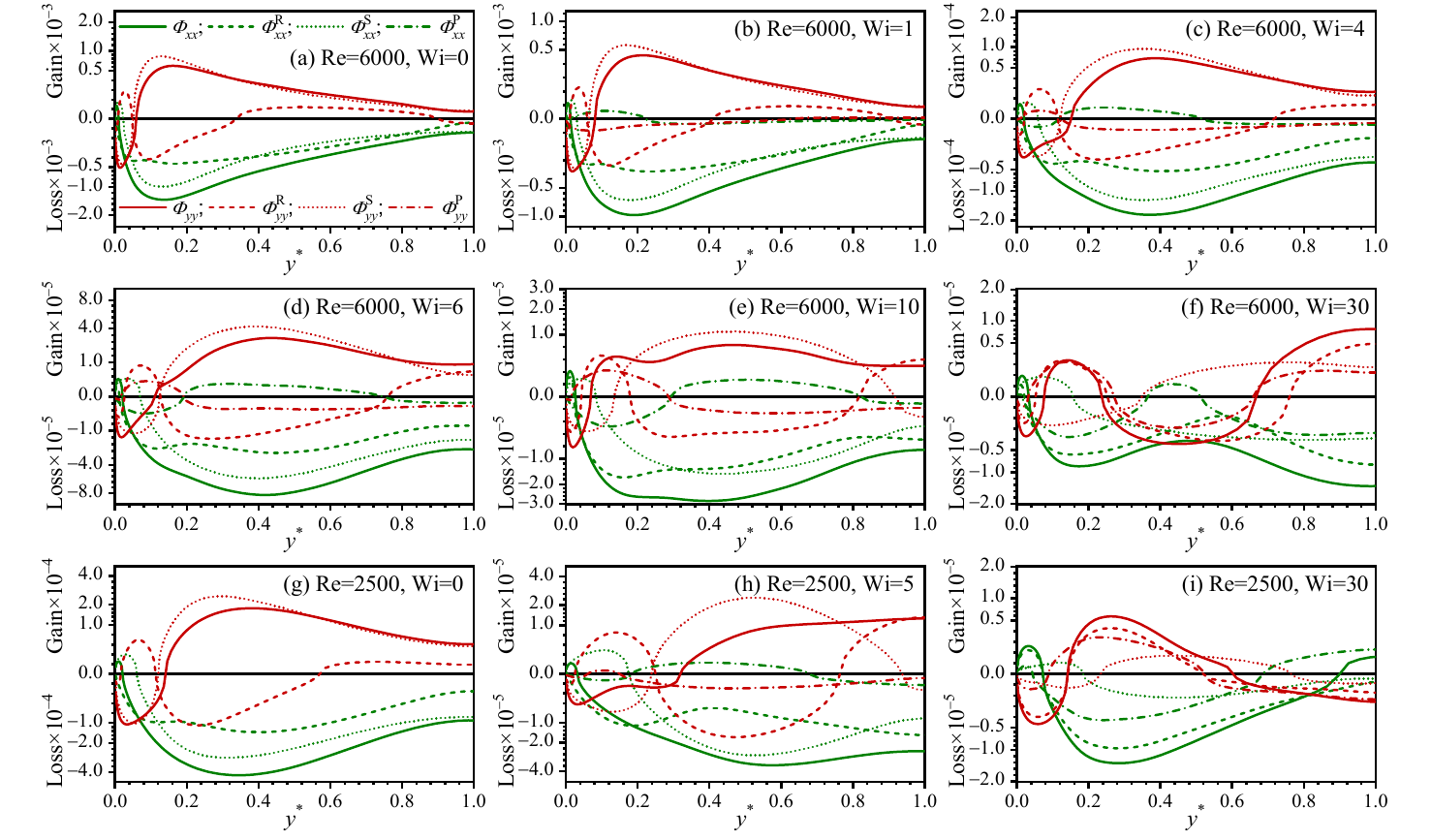}
\caption{\label{fig4} Profiles of total pressure distribution effect on $\overline{{{u'}^ {*}}{{u'}^{*}}}$, $\overline{{{v'}^ {*}}{{v'}^{*}}}$ and the divided rapid, slow and elastic contributions at: (a) Re = 6000, Wi = 0 (Newtonian IT); (b) Re = 6000, Wi = 1; (c) Re = 6000, Wi = 4; (d) Re = 6000, Wi = 6; (e) Re = 6000, Wi = 10 (MDR); (f) Re = 6000, Wi = 30 (EIT); (g) Re = 2500, Wi = 0 (Newtonian IT); (h) Re = 2500, Wi = 5 (before relaminarization); (i) Re = 2500, Wi = 30 (EIT).}
\end{figure}

\subsection{Re-picturing viscoelastic drag-reducing turbulence}
\begin{figure}
\centering
\includegraphics[width=0.6\textwidth]{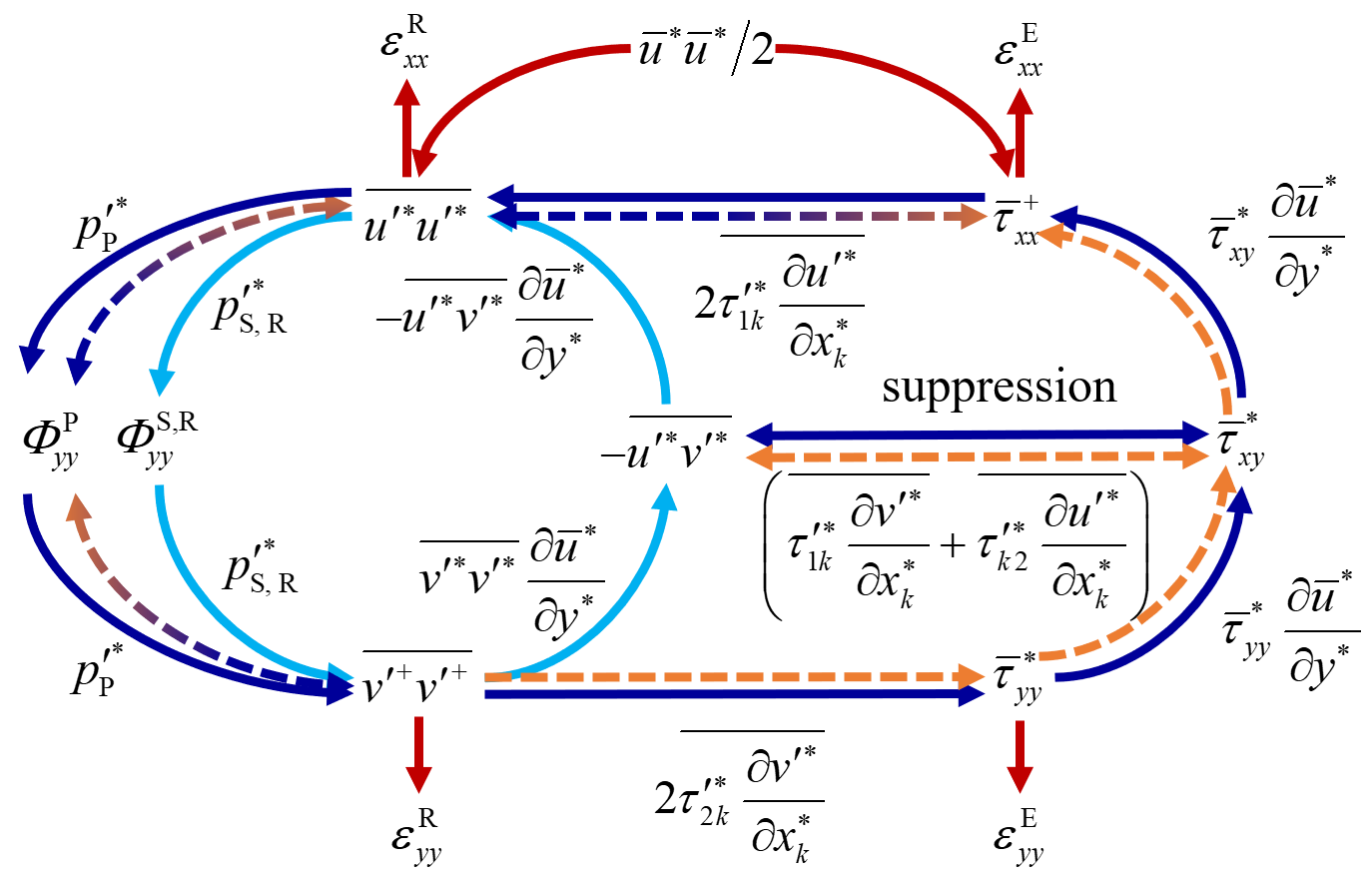}

\caption{\label{fig5} Energy picture of viscoelastic DRT. \textcolor{cyan}{\bf{---}} and \textcolor{blue}{\bf{---}}: SSP cycles in IT and EIT. The \textcolor{orange}{\bf{---}}: the bypass of energy process caused by the passive response of polymers on IT. \textcolor{orange}{\bf{-}} \textcolor{blue}{\bf{-}}: the path change from passive response on IT SSP to actively participating in EIT SSP with the increase of Wi. \textcolor{red}{\bf{---}}: energy source and dissipation.}
\end{figure}

Based on the analysis, we re-picture the energy transfer process for viscoelastic DRT from IT to EIT by combing TKE and stress budgets as shown in Figure {\ref{fig5}}. The cycles of SSP for IT and EIT are illustrated by the cyan paths and the blue paths, repectively. The energy process in DRT from onset of DR to MDR, is illustrated by the dashed paths colored gradient from orange to blue. Under LDR conditions, polymers passively respond to the SSP of IT. TKE is partially transformed into EE through the interaction between polymers and turbulence in both the streamwise and wall-normal directions. NESS is also formed due to wall-normal polymer extension. NESS absorbs energy from mean motion and turns into streamwise EE. The above process is a passive response to the SSP of IT, which cannot support but intervent its SSP. Pressure fluctuations induced by polymers distribute TKE from the wall-normal direction to other directions such as the streamwise which is opposing with that of inertial pressure distribution although with almost neglibile magnitude. Moreover, there is an orange path showing the interaction between polymer and turbulence, which inhibits both RSS and NESS and is another important factor for polymers to intervent with the SSP of IT. When Wi is high enough, the energy process of DRT gradually transits to that of EIT state. In this process, the SSP of IT weakens and SSP cycle of EIT dominates the flow replacing that of IT.

With this energy picture in mind, the different DR phenomena at low Re and moderate Re can be well unified. As shown in Figures 3 and 4, the two important links  related with SSP cycle of EIT discribed by $G_{xx}$ and $\Phi^P_{yy}$ are not formed untile $Wi = 5$ for $\rm{Re}=2500$, but can be well hold even at $\rm{ Wi}=4$ for $\rm{Re}=6000$. Further increasing Wi, at $\rm{Re}=2500$ the flow is laminarized since the weaker IT disappears before EIT is excited, whereas, at $\rm{Re}=6000$  the flow finally enters EIT without a period of relaminarization as EIT can be well sustained before IT disappears.

We can also easily understand the reason why $u'$ lags behind $v'$ in DRT observed in Figure 2. In IT, DRT and EIT, the formation of $u'$ are all inseparable with $v'$ but with different routine. In IT, the energy absorbed by RSS sustains the formation of $u'$. The formed TKE in streamwise is then redistributed to $v'$ directly through the pressure redistribution effect. The formation $u'$ and the energy redistribution to $v'$ are direct processes and rapid terms. However, in DRT and EIT regime, polymers modulate even change the formation mechanism of $u'$, as illustrated by the orange path and blue path in Figure {\ref{fig5}}. Different from that in IT, $u'$ is mainly derived from the conversion of streamwise EE in EIT regime, and the generation of streamwise EE is from the wall-normal EE caused by $v'$. Due to the polymers memory effect, the formation and development of this process are indirect and takes a certain time, related to the fluid relaxation time, which results in the phenomenon $u'$ lags behind $v'$. For DRT, with the increase of Wi, the IT related SSP gradually weakens, while the EIT related SSP enhances and the time memory effect increases, resulting in the phenomenon that the lag distance increases.

\section{Conclusions }
In summary, we have re-pictured viscoelastic turbulent channel flow, especially its TKE transfer process, based on statistical and budget analysis throughout a wide range of flow state covering IT, LDR, HDR, MDR to EIT and relaminarized flow regime. The evolutionary comparison of $P_k$, $P_e$ and $G$ with Wi provide quantitative evidence that the nature of DRT gradually changes from IT to EIT. The statistical results of $C_{uv}$ among IT, DRT and EIT are in sharp contrast, which further implies that there exists a SSP different from IT in DRT. For energy and stress budget, polymers do have the passive feedback effect of modulating and interfering with the SSP of IT, but more importantly, when Wi exceeds a certain critical value long before the flow enters MDR state, polymers can actively excite EIT-related SSP. The two obvious characteristics of this active behavior are that polymers pump energy from mean motion to induce streamwise turbulent fluctuations, and that polymers can induce the pressure distribution to provide energy for the formation of wall-normal turbulent fluctuations. These evidences fully suggest that the nature of DRT experienced the transition from pure IT, IT modulated by polymers, participation of EIT and finally maybe pure EIT. The different drag reduction phenomena at low (Re=2500) and moderate (Re=6000, 20000) Re are unified through the proposed energy picture. The reason why the relaminarized phenomenon doesn't occur at moderate Re is that EIT dynamics gets involved in the DRT earlier, so that the turbulence can be maintained and feeded by elasticity instability. The proposed energy picture can well illustrate why the existing DRT models cannot well predict HDR and MDR flows since EIT dynamics are not involved, also provides a theoretical fundament for DRT modelling.


\bibliographystyle{alpha}

\end{document}